\documentclass[aps,twocolumn,preprintnumbers,amsmath,amssymb]{revtex4-2}
\usepackage[english]{babel}
\usepackage{graphicx}
\usepackage[normalem]{ulem}
\usepackage[svgnames]{xcolor}
\usepackage{bm}
\usepackage{multirow}
\usepackage{float}
\usepackage{titlesec}
\usepackage[columnwise]{lineno}
\usepackage{float}
\usepackage{titlesec}
\usepackage{gensymb}
\usepackage{textcomp}
\usepackage{url}

\begin{document}

\title{Incipient modulated phase in Sr$_{1-x}$Ca$_{x}$TiO$_3$}
%\orcid{0000-0002-6457-691X}
\author{Beno\^{\i}t Fauqu\'e$^{1}$}
 \email{benoit.fauque@espci.fr}
\author{Daniel A. Chaney $^{2}$}
\author{Philippe Bourges$^{3}$}
\author{St\'ephane Raymond$^{4}$}
\author{Arno Hiess$^{5,6}$}
\author{Paul Steffens$^{5}$}
\author{Benoît Baptiste$^{7}$}
\author{Luigi Paolasini $^{2}$}
\author{Alexe\"{i} Bosak $^{2}$}
\author{Kamran Behnia $^{1}$}
\author{Yasuhide Tomioka$^{8}$}

%\orcid{0000-0002-6457-691X}

\affiliation{
$^{1}$ Laboratoire de Physique et d'Etude de Mat\'{e}riaux (CNRS)\\ ESPCI Paris, Université PSL, 75005 Paris, France
$^{2}$ European Synchrotron Radiation Facility, 71 avenue des Martyrs, 38000 Grenoble, France \\
$^{3}$ Laboratoire L\'eon Brillouin, CEA-CNRS, Universit\'e Paris-Saclay, CEA Saclay, 91191 Gif-sur-Yvette, France\\
$^{4}$ Univ. Grenoble Alpes, CEA, IRIG, MEM, MDN, 38000 Grenoble, France\\
$^{5}$ Institut Laue-Langevin, 71 Avenue des Martyrs, 38042 Grenoble Cedex 9, France\\
$^{6}$ European Spallation Source ERIC, P.O. Box 176, 22100 Lund, Sweden
$^{7}$ IMPMC-Sorbonne Université and CNRS, 4, place Jussieu, 75005 Paris, France\\
$^{8}$ National Institute of Advanced Industrial Science and Technology (AIST), Tsukuba, 305-8565, Japan\\
}

\date{\today}
\begin{abstract}
Nanometer-scale modulations can spontaneously emerge in complex materials when multiple degrees of freedom interact. Here we demonstrate that ferroelectric Sr$_{1-x}$Ca$_x$TiO$_3$ lies in close proximity to an incipient structurally modulated phase. Using inelastic neutron and X-ray scattering, we show that upon cooling, dipolar fluctuations strongly couple to and soften the $c_{44}$ transverse acoustic mode. We identify the wavevector at which this softening is maximal, thereby defining the characteristic length scale of the modulation. Calcium substitution enhances both the amplitude and the wavevector of the softening by strengthening the ferroelectric and antiferrodistortive instabilities. Our results demonstrate that nonlinear flexoelectric phonon coupling tends to stabilize a modulated state that cooperates with, rather than competes against, the other lattice instabilities in SrTiO$_3$.
\end{abstract}

\maketitle

Competing interactions may lead to modulated phases in condensed matter, characterized by periodic spatial variations of the relevant order parameter \cite{Seul1997}. The modulation period is set by the relative strengths of the active degrees of freedom and can be tuned by temperature or external fields. Such states, found in manganites \cite{Dagotto2005,Milward2005}, high-temperature cuprates \cite{Keimer2015}, and two-dimensional electron gases under strong magnetic fields \cite{Lilly1999}, give rise to rich phase diagrams and novel functionalities. Understanding how these modulated phases interact with other lattice or electronic instabilities is therefore crucial for both fundamental insight and materials design. Recently, SrTiO$_3$ has been identified as a new member of this class of nanoscale-modulated materials \cite{Fauque2022,Gian2023,orenstein2025,zhang2025,Kaplan2025}.

\begin{figure*}[ht!]
%\centering
\centering
\makebox{
\includegraphics[width=1.0\textwidth]{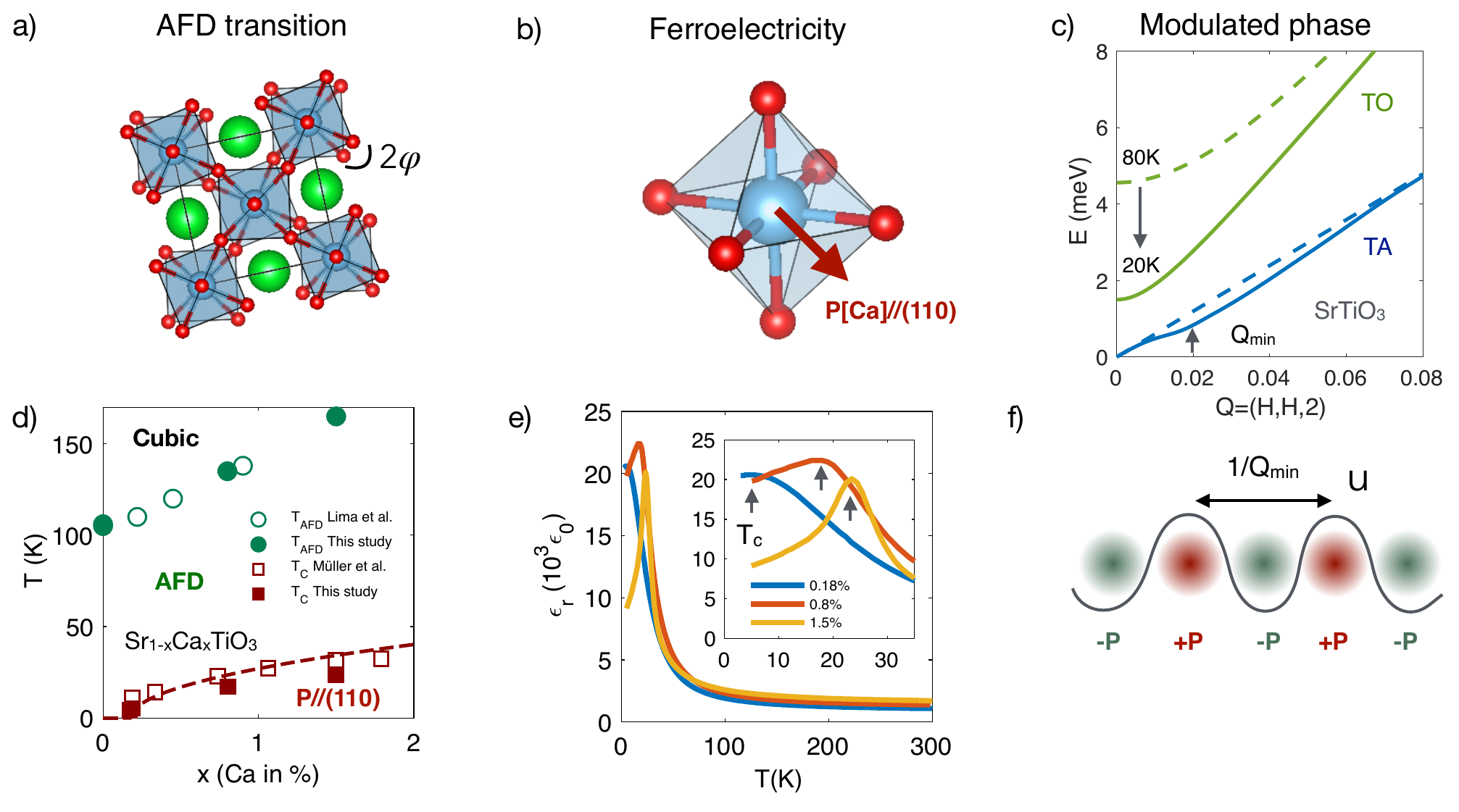}}
\caption{{\textbf{Lattice instabilities in Sr$_{1-x}$Ca$_{x}$TiO$_3$}} (a) Antiferrodistortive (AFD) transition at $T_{\mathrm{AFD}}$.
(b) Polar distortion: Ca$^{2+}$ substitution stabilizes ferroelectricity along the [110] direction.
(c) Schematic of TO–TA coupling in Sr$_{1-x}$Ca$_x$TiO$_3$: softening of the transverse-optical (TO) mode induces a concomitant softening of the transverse-acoustic (TA, $c_{44}$) branch, which is maximal at $Q_{\mathrm{min}}$, signaling a tendency toward a modulated phase.
(d) Phase diagram of Sr$_{1-x}$Ca$_x$TiO$_3$: $T_{\mathrm{AFD}}$ and Curie temperature $T_C$ as a function of Ca concentration, compared with previous studies \cite{Lima2015,Bednorz1984}.
(e) Dielectric constant as a function of temperature for three Ca concentrations; the inset highlights the low-temperature region below 35 K. (f) Real-space illustration of TO–TA coupling: dipolar fluctuations are accompanied by a transverse strain modulation with wavelength $\sim 1/Q{\mathrm{min}} \simeq 15$ nm in pristine SrTiO$_3$.}.
\label{FigIntro}
%\end{center}
\end{figure*}

%STO : AFD + Ferroelectric  : 
SrTiO$_3$ is a paradigmatic system for exploring novel structural and electronic phases \cite{MullerBook:1981,CollignonRev2019}, largely due to its proximity to multiple lattice instabilities. Two are well-established: the anti-ferro-distortive (AFD) transition, where TiO$_6$ octahedra rotate in an anti-phase manner (Fig.~\ref{FigIntro}a), and the quantum-paraelectric regime, where dipolar fluctuations grow upon cooling yet never condense into long-range ferroelectric order \cite{Muller1979}.

Despite decades of study, the quantum-paraelectric phase of SrTiO$_3$ remains unsettled. It has been described either as an incipient ferroelectric state suppressed by zero-point fluctuations \cite{Schneider1976,Muller1979,Shin2021,esswein2021}, or as a regime where the soft transverse-optical (TO) mode hybridizes with an acoustic branch \cite{Muller1991,Bussmann2007,Rowley2014,Coak12707}. Coupling between the TO mode and the transverse-acoustic (TA) $c_{44}$ branch has been documented for decades \cite{Yamada1969,Vacher1992,Courtens1993,Balashova1996,Hehlen1996,Hehlen1998,Delaire2020}, but only recently did Fauqué \emph{et al.} \cite{Fauque2022} show that the TA softening is strongly wavevector-dependent, with a pronounced anomaly along the (110) direction at $Q_{\mathrm{min}} \simeq 0.02$ (r.l.u.), sketched on (Fig.~\ref{FigIntro}c). This identifies a modulated state in which polarization fluctuations are accompanied by a transverse strain modulation with a characteristic wavelength of $\sim 15$ nm (Fig.~\ref{FigIntro}f). Recent complementary observations have reinforced this picture including ultrafast X-ray diffraction revealing this polar–acoustic collective mode \cite{orenstein2025} and STEM imaging of fluctuating polar nanoscale domains displaying similar length scales \cite{zhang2025}.

This nanoscale modulation is governed purely by lattice degrees of freedom and nonlinear phonon interactions. Within Landau–Ginzburg–Devonshire (LGD) theory \cite{Morozovska2017,Gian2023}, such a state can emerge at the mean-field level when the softening TO mode approaches the TA branch. The flexoelectric coupling, which links polarization to strain gradients, can drive the TA mode unstable at a finite wavevector and, if sufficiently strong, stabilize a static incommensurate structure \cite{Morozovska2017}. However, once thermal and quantum fluctuations are included this ordered state becomes fragile, giving way to a competition between homogeneous ferroelectricity and a melted, fluctuating version of the modulated phase \cite{Gian2023}.

Here we show that, in Sr$_{1-x}$Ca$_x$TiO$_3$, the melted modulated phase does not compete with ferroelectricity but instead reinforces it. Both the magnitude and the characteristic wavevector of the TA softening increase with Ca substitution, which simultaneously enhances the ferroelectric and AFD instabilities. At the highest dopings, the TA branch develops a waterfall-like dispersion that lies beyond the mean-field flexoelectric theory. Together, these results demonstrate that SrTiO$_3$ hosts an incipient modulated phase that is intrinsically intertwined with its other lattice instabilities, providing a microscopic foundation for its nanoscale structural modulation.

Figure \ref{FigIntro}d presents the phase diagram of Sr$_{1-x}$Ca$_x$TiO$_3$, showing the evolution of the ferroelectric ($T_C$, red) and antiferrodistortive ($T_\text{AFD}$, green) transitions with Ca substitution. Both $T_C$ \cite{Bednorz1984} and $T_\text{AFD}$ \cite{Lima2015} increase as Ca content rises. The temperature dependence of the dielectric constant for three single crystals ($x = 0.18\%$, $0.8\%$, $1.5\%$) is shown in Fig.~\ref{FigIntro}e, confirming that ferroelectricity emerges above a critical Ca concentration $x_c \simeq 0.18\%$ \cite{Bednorz1984} along the (110) direction (Fig.~\ref{FigIntro}b). Sample preparation and characterization details are provided in \cite{SM}.

\begin{figure*}[ht!]
\centering
\makebox{\includegraphics[width=1.0\textwidth]{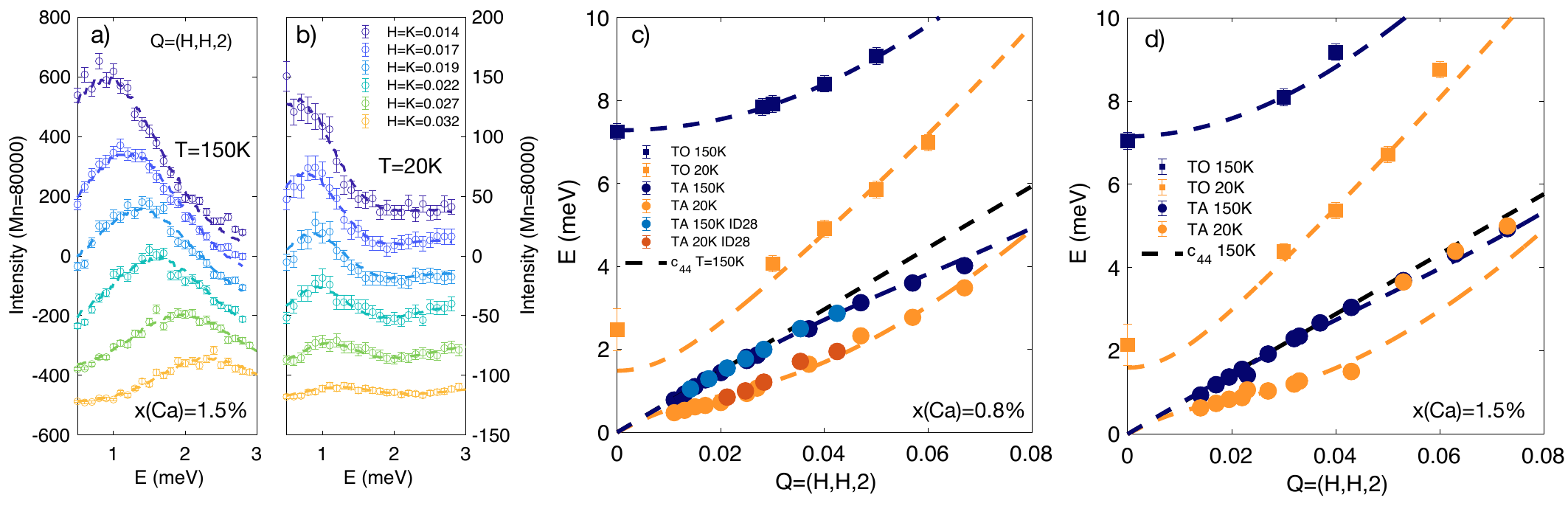}}
\caption{{\textbf{TA softening in  Sr$_{1-x}$Ca$_x$TiO$_3$ : }} 
(a–b) Energy scans at $\mathbf{Q} = (H,H,2)$ for $H = K = 0.014$–0.032 at (a) $T = 150$~K and (b) $T = 20$~K for $x$ = 1.5 $\%$. Fits including a convolution with the experimental resolution are shown in dotted lines (see \cite{SM})(c) Dispersion of the TO (squares) and TA (circles) modes for $x = 0.8\%$ at $T = 150$~K (dark blue: INS; light blue: IXS) and $T = 20$~K (orange: INS; red: IXS). (d) Same as (c) for $x = 1.5\%$. Both dispersions are fitted using the mean-field Landau–Ginzburg–Devonshire solution \cite{Morozovska2017} (see text). At the highest doping, the model fails to capture the abrupt change of the TA dispersion around $H = K = 0.04$.}
\label{Fig1p5}
%\end{center}
\end{figure*}

The low-energy phonon spectrum was measured using the cold triple-axis spectrometers IN12 and THALES at the Institut Laue-Langevin, Grenoble, with samples mounted in the (H,H,L) scattering plane in the (002) Brillouin zone. High-resolution configurations probed the transverse acoustic (TA) $c_{44}$ branch near the Bragg peak, while standard settings accessed larger wavevectors ($H \gtrsim 0.03$). Complementary inelastic X-ray scattering (IXS) measurements were performed on ID28 at ESRF. Experimental details and data analysis procedures are given in Sections B and C of \cite{SM}.

\begin{figure}
\includegraphics[width=0.5\textwidth]{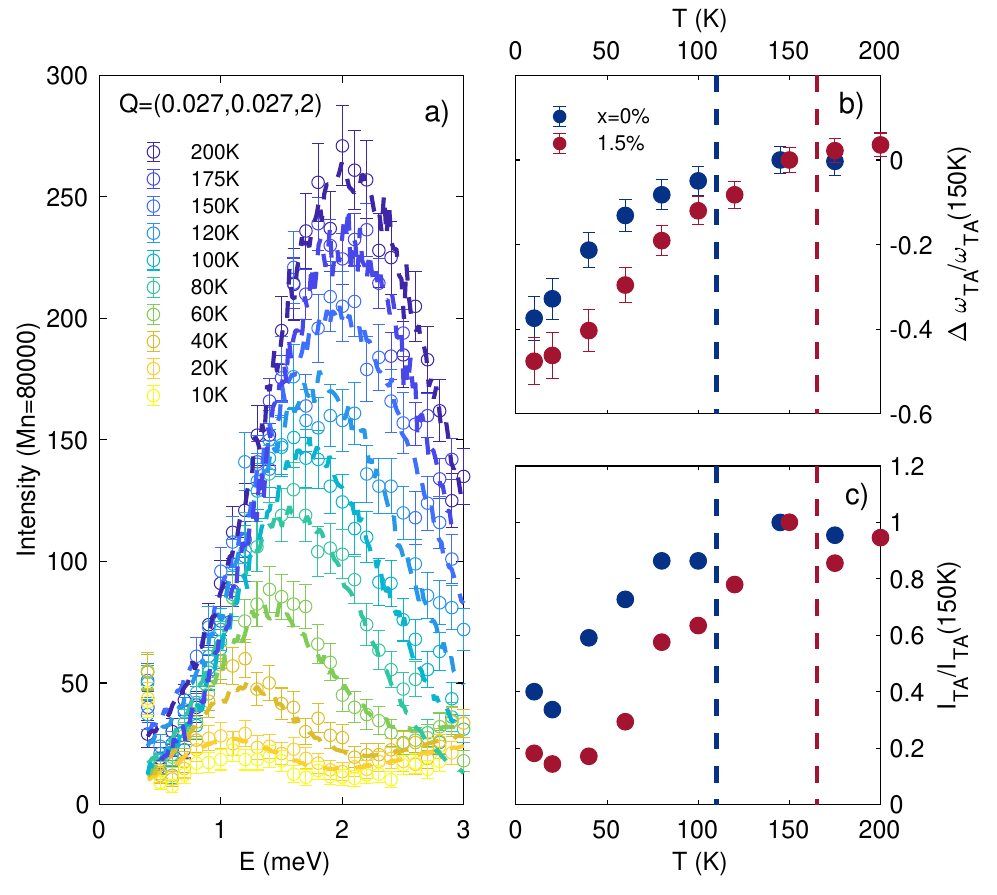}
\caption{{\textbf{Temperature dependence of the TA softening in Sr$_{1-x}$Ca$_x$TiO$_3$ : } Energy scans at $\mathbf{Q} = (0.027,0.027,2)$ in the Ca-doped sample ($x = 1.5\%$) from $T = 10$–200~K (high-resolution mode). Fits including a convolution with the experimental resolution are shown in dotted lines (see \cite{SM}) (b–c) Temperature dependence of the TA mode: (b) energy and (c) intensity for $x$=0 (dark blue) at $H$=$K$=0.017 and 1.5$\%$ (dark red) at $H$=$K$=0.027. The dashed vertical line indicates the structural transition $T_{\text{AFD}}$. The TA mode softening begins below $T_{\text{AFD}}$.}}
\label{FigTemp0p025}
\end{figure}

Figure \ref{Fig1p5}a,b show energy scans for the 1.5$\%$ Ca-doped sample at ${\bf Q} = (H,H,2)$, taken for $H = 0.017$ to $0.035$ at 150 K and 20 K, respectively. At high temperature, the TA mode energy increases with $H$, while at low temperature the mode remains nearly constant across the same momentum range, revealing a pronounced softening. To extract quantitative dispersions, the energy scans were fitted with a damped harmonic oscillator (DHO) model convolved with the experimental resolution (see \cite{SM}). The resulting TO and TA dispersions for $x = 0.8\%$ and $x = 1.5\%$ are shown in Fig.~\ref{Fig1p5}c,d. At 150 K, the TA mode exhibits a quasi-linear dispersion, in agreement with the $c_{44}$ elastic constant measured by ultrasound \cite{Carpenter2007}. At low temperatures, the TA mode deviates strongly from linearity, forming a plateau at $H = 0.025$ for $x = 0.8\%$ and at $H \approx 0.035$ for $x = 1.5\%$. At the highest doping, the dispersion shows a step-like behavior, reminiscent of the "waterfall" dispersion observed in the TO mode of relaxor perovskites \cite{Gehring2000,Gehring2001,Hlinka2003}.

The temperature dependence of the TA mode at $H \approx 0.027$ is shown in Fig.~\ref{FigTemp0p025}. Upon cooling, the mode softens by roughly 50$\%$, accompanied by a sixfold decrease in intensity. The softening commences just below the AFD transition ($T_{\text{AFD}} \simeq 165$ K at $x = 1.5\%$), in agreement with previous studies of pure SrTiO$_3$ \cite{Fauque2022}.

Independent confirmation of the presence of the TA softening across multiple Brillouin zones is provided by IXS measurements for $x = 0.8\%$, performed around $\bm{Q} = (4,4,0)$. Despite differences in energy resolution (IXS $\sim 1.5$ meV vs. INS $\sim 0.4$ meV) and Q-space coverage, the extracted dispersions are in excellent agreement, as shown in Fig.~\ref{Fig1p5}c, confirming the robustness of the neutron results. We note that INS allowed us to probe the TA mode slightly closer to the Bragg peak when compared to IXS. This is likely due to the superior energy resolution of the INS technique.

\begin{figure*}[ht!]
\centering
\makebox{\includegraphics[width=1.0\textwidth]{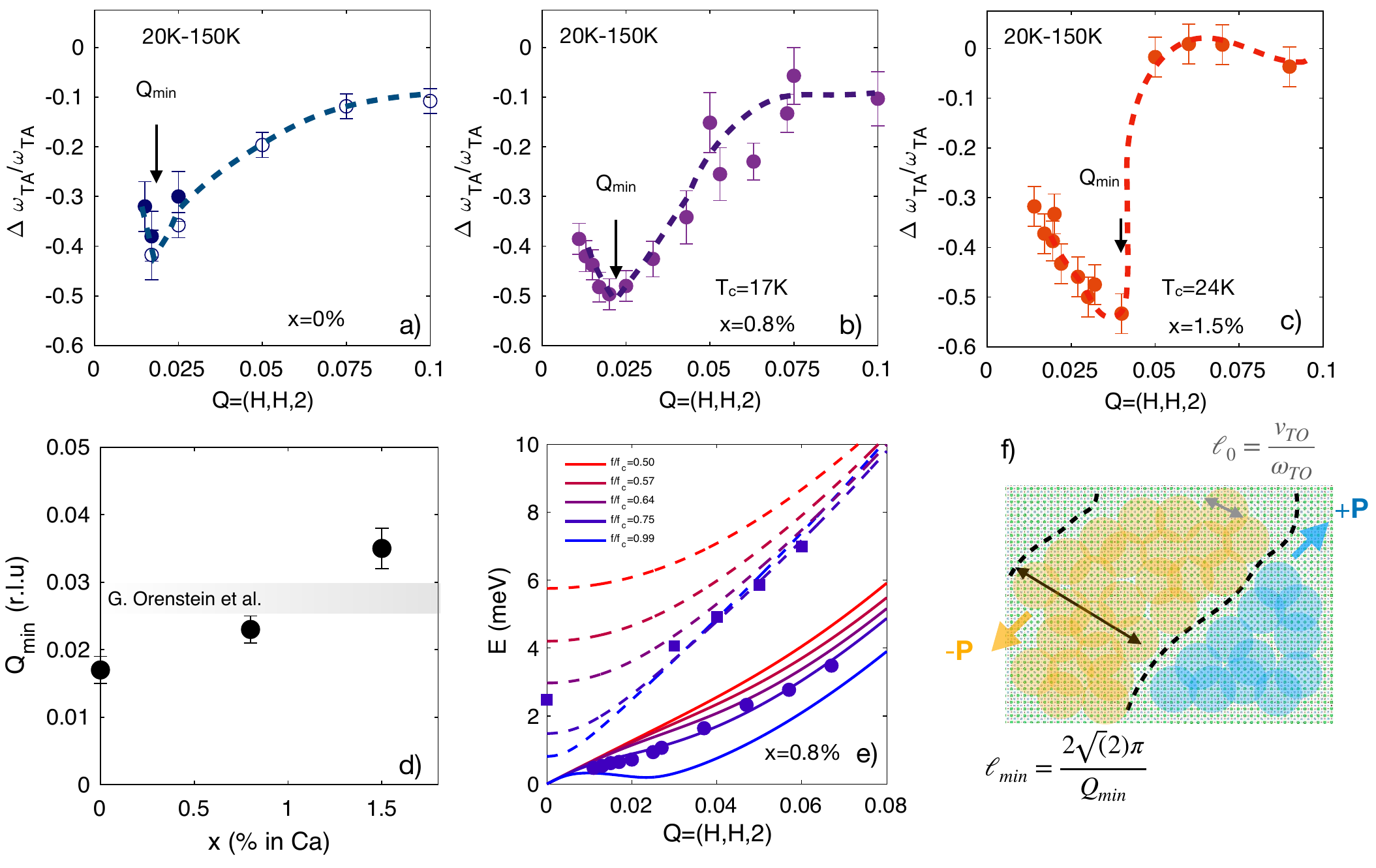}}
\caption{{\textbf{Doping evolution of the TA softening in Sr$_{1-x}$Ca$_x$TiO$_3$: } Q-dependence of the TA softening between 20~K and 150~K, $\frac{\Delta \omega_{TA}}{\omega_{TA}} = [\omega_{TA}(20~\mathrm{K}) - \omega_{TA}(150~\mathrm{K})]/\omega_{TA}(150~\mathrm{K})$, for (a) $x = 0$, (b) $x = 0.8\%$, and (c) $x = 1.5\%$ in Sr$_{1-x}$Ca$_x$TiO$_3$. Dashed lines are a guide to the eye. Closed circles are results from this work; open circles are taken from Ref.~\cite{Fauque2022}. (d) Doping evolution of $Q_{\rm min}$. The gray band indicates the wavevector where the TO–TA coupling was found to be maximal in Ref.~\cite{orenstein2025}. (e) Dispersion of the TO (dotted lines) and TA (solid lines) modes in the presence of a flexoelectric coupling $f$ \cite{Morozovska2017}. As the TO mode softens, $f$ approaches the critical value $f_c$, above which an incommensurate phase is stabilized. Closed squares and circles denote the TO and TA modes, respectively, for $x = 0.8\%$. (f) Real-space schematic illustrating the two characteristic length scales in Sr$_{1-x}$Ca$_x$TiO$_3$: around a Ca dopant, a ferroelectric moment forms over a length $\ell_0$, which clusters with other dipoles over a longer scale $\ell_{\rm min} = 2\sqrt{2}\pi/Q_{\rm min}$.}}
\label{FigDoping}
%\end{center}
\end{figure*}

Our measurements reveal that Ca doping not only steepens the TA softening in Sr$_{1-x}$Ca$_x$TiO$_3$ (Fig.~\ref{FigDoping}a-c), but also shifts the minimum to higher $Q$ (Fig.~\ref{FigDoping}d). Even at the highest doping, the softening remains incomplete, and no static incommensurate order is observed (see \cite{SM} Section B), indicating that the system develops a dynamically modulated lattice instability rather than a static modulation. Remarkably, the $Q_{min}$ in doped samples coincides with the wavevector of maximal TO–TA coupling under terahertz excitation in pure SrTiO$_3$ \cite{orenstein2025}, suggesting that ferroelectricity, whether induced by doping or light \cite{li2019Thz,nova2019,Fechner2024}, shifts $Q_{\text{min}}$ to higher values.

To fit the TA and TO dispersions for $x = 0.8\%$ and $x = 1.5\%$, we used a mean-field solution of the Landau–Ginzburg–Devonshire (LGD) model including flexoelectric coupling \cite{Morozovska2016,Morozovska2017}, shown as dotted lines in Fig.~\ref{Fig1p5}c,d. In this framework, as the TO mode softens and approaches the TA branch, the flexoelectric coupling $f$ induces a softening of the TA mode that is maximal at a finite wavevector. When $f$ approaches a critical value $f_c$, the TA branch develops a minimum that eventually reaches zero at $f=f_c$, signaling the onset of an incommensurate, spatially modulated phase. Figure~\ref{Fig1p5}e illustrates the evolution of both modes for different ratios $f/f_c$ and compare with the TA and TO dispersions for $x = 0.8\%$.

To limit the number of fitting parameters, we fixed the elastic constants to values obtained from ultrasound measurements \cite{Carpenter2007} and considered only static flexoelectricity. Only three parameters were adjusted to reproduce the TA dispersion (see \cite{SM}, Sec.~D). Using $f \simeq 6$~V, comparable to experimentally reported values \cite{Mizzi2022}, the model quantitatively reproduces both TO and TA dispersions for $x = 0.8\%$, corresponding to $f/f_c \simeq 0.8$. In contrast, it fails to capture the waterfall-like dispersion observed at $x = 1.5\%$, suggesting that a more sophisticated approach, including a $q$-dependent flexoelectric coefficient or self-consistent phonon approximation is required.

Contrary to the model proposed by \cite{Gian2023}, where ferroelectricity and modulated phase instabilities compete, we observe their simultaneous and cooperative presence. One identifiable source of this cooperation is the splitting of the TO mode below $T_{\text{AFD}}$  \cite{Uwe1976}. The tetragonal distortion splits the polar mode into $A_{2u}$ and $E_u$ branches, with the splitting proportional to the octahedral tilt angle $\varphi$ \cite{Yamanaka2000,Hehlen2013}. Ca doping increases $\varphi$ \cite{Geneste2008}, enhancing the splitting of the polar soft mode, which in turn favors the emergence of a modulated phase. Interestingly, we show that the TA softening begins once $T < T_{\mathrm{AFD}}$, see Fig. \ref{FigTemp0p025}b and c), providing further evidence that both the AFD distortion and the modulated phase cooperate.

These momentum-space instabilities point to multiscale ferroelectric structures, such as the one illustrated on Fig.~\ref{FigDoping}f). In highly polarizable Sr$_{1-x}$Ca$_x$TiO$_3$, local dipoles form around Ca atoms, breaking inversion symmetry \cite{Vugmeister1990,Samara_2003}. As the Ca concentration increases, these dipoles percolate into a macroscopic ferroelectric state \cite{Bianchi1995,Samara_2003}. Our work reveals a second, larger length scale, $\ell_{min} \sim 10\;\mbox{nm}$, associated with dynamically fluctuating domains. The coexistence of short-range ferroelectric regions and extended modulations demonstrates that the ground state is inherently multiscale, combining local ferroelectric order with a dynamically modulated lattice texture.

In conclusion, we show the quantum paraelectric regime of SrTiO$_3$ lies at the intersection of three intertwined lattice instabilities: ferroelectricity, AFD transition, and modulated phase. This finding is essential for understanding its unusual lattice properties, ranging from electric permittivity \cite{Coak12707,Rowley2014,Li2025} and thermal conductivity \cite{Martelli2018} to low-frequency ultrasound \cite{Kustov2020} and light-induced phases \cite{nova2019,Li2019,Fechner2024,orenstein2025}, and may also underline the enhancement of superconductivity in ferroelectric superconductors \cite{Rischau2017,Tomioka2019,Ahadi2019,Rischau2022,Tomioka2022,Zhang2025b}. Controlling this dynamical modulation via strain, electric fields, or carrier doping provides a promising route for engineering emergent ferroic behavior in quantum paraelectrics and their interfaces.

 \section*{Acknowledgements}
 We thanks A. Bussmann-Holder, G.G Guzm\'an-Verri, P. Littlewood, A. N. Morozovska, E. Salje, A. Subedi and N. Spaldin for useful discussions. This work was supported by Jeunes Equipes de l$'$Institut de Physique du Coll\`ege de France and by a grant attributed by the Ile de France regional council. We acknowledge the European Synchrotron Radiation Facility for provision of beamtime under the proposal HC-5544. This work was also supported by Japan Society for the Promotion of Science (JSPS) KAKENHI Grant Numbers 23H01135 and 23K25832. 

 \section*{Data availability} 
 INS data are available at \url{https://doi.ill.fr/10.5291/ILL−DATA.7−01−572}, \url{https://doi.ill.fr/10.5291/ILL-DATA.CRG-2926} and \url{https://doi.ill.fr/10.5291/ILL−DATA.7−01−606}. The rest of the data that support the findings of this study is available from the corresponding authors upon request.

\end{document}